\documentstyle[12pt]{article}
 \textwidth
 16.4cm
 \oddsidemargin
 2.5cm
 \advance\oddsidemargin
 by
 -1in
 \evensidemargin
 0.0cm
 \advance\evensidemargin
 by
 -1in
 \marginparwidth
 1.9cm
 \marginparsep
 0.4cm
 \marginparpush
 0.4cm
 \topmargin
 -1.5cm
 \advance\topmargin
 by
 -0.0in
 \textheight
 23.0cm
 \makeindex

 \newcommand{\doublespace}
 {
 \renewcommand{\baselinestretch}
 {1.6}
 \large\normalsize}
 \newcommand\la{\langle}
 \newcommand\ra{\rangle}
 \newcommand\noi{\noindent}
 \newcommand\beq{\begin{equation}}
 \newcommand\eeq{\end{equation}}
 \newcommand\beqn{\begin{eqnarray}}
 \newcommand\eeqn{\end{eqnarray}}

\begin{document}

\vspace*{3cm}



\centerline{\Large
\bf
Challenges
of
Nuclear Shadowing
in
DIS}

\vspace{.5cm}

\begin{center}
 {\large
B.Z.~Kopeliovich$^{1,3}$,
J.~Raufeisen$^{2}$
and
A.V.~Tarasov$^{2,3}$}\\
\medskip

{\sl
$^1$
Max-Planck
Institut
f\"ur
Kernphysik,
Postfach
103980,
69029
Heidelberg,
Germany}\\

{\sl $^2$Institut
f\"ur
Theoretische
Physik
der
Universit\"at,
Philosophenweg
19,
\\
69120
Heidelberg,
Germany}\\

{\sl
$^3$
Joint
Institute
for
Nuclear
Research,
Dubna,
141980
Moscow
Region,
Russia}

\end{center}

\vspace{.5cm}
\begin{abstract}

Nuclear shadowing in DIS at moderately small $x$
is
suppressed by the nuclear formfactor and depends on
the effective
mass of a hadronic fluctuation of the virtual photon.  
We propose a solution to the problems (i) of how to combine
a definite transverse size of the
fluctuation
with a definite effective mass, and (ii) of 
how to include the nuclear formfactor in the
higher
multiple scattering terms.
Comparison of the numerical results
with
known approximations shows a substantial
difference.

\end{abstract}

\bigskip

\newpage

\doublespace

\noi
{\large\bf
1. Introduction}
\medskip

Shadowing in deep-inelastic
scattering
(DIS)
off nuclei is a hot topic for the last
two
decades.
In the infinite momentum frame of
the
nucleus it can be interpreted 
as a result of parton fusion leading to
a
diminishing
parton density at low Bjorken
$x$
\cite{kancheli} - \cite{q}.
A more intuitive picture arises in the rest
frame
of
the nucleus where the same phenomenon looks like
nuclear
shadowing
of hadronic fluctuations of the virtual
photon
\cite{lp} - \cite{kp}.
To crystallize the problem and its solution we
restrict
ourselves
in this paper to only quark-antiquark fluctuations of
the
photon,
neglecting those higher Fock components which contain gluons
and $q\bar q$ pairs from the sea. The lifetime of the $q\bar
q$
fluctuation
(called coherence time) is
given
by
\beq
t_c
=
\frac{2\nu}{Q^2+M^2}
\label{1.1}
\eeq
\noi
where $\nu$ is the photon energy, $Q^2$ its virtuality and
$M$
is
the effective mass of the $q\bar
q$
pair.

Provided that the coherence time is much longer than the
nuclear
radius,
$l_c \gg R_A$, the total cross section on
a
nucleus
reads
\cite{zkl},
\beqn
\sigma^{\gamma^*A}_{tot}(x,Q^2)
&=&
2\,\int
d^2b
\int
d^2
r\,G_{\gamma^*}(Q^2,r)\,
\left\{1-\exp\left[-{1\over
2}\sigma(r)
T(b)\right]\right\}\nonumber\\
&\equiv&
2\,\int
d^2b\
\left\{1-
\left\la
\exp\left[-{1\over
2}\sigma(r)
T(b)\right]\right\ra\right\}\ .
\label{1.2}\label{eik}
\eeqn
\noi
Here $G_{\gamma^*}(Q^2,r)$ characterizes the
probability
for
the photon to develop a $q\bar q$
fluctuation
with
transverse separation $r$. The condition
$t_c\gg
R_A$
insures that the $r$ does not vary
during
propagation
through the nucleus (Lorentz
time
dilation).
Then the $q\bar q$ pair with a definite
transverse
separation
is an eigenstate of the interaction with the
eigenvalue
of
the total cross
section
$\sigma(r)$.
Therefore, one
can
apply
the eikonal expression (\ref{1.2}) for the interaction with
the
nucleus.
The nuclear thickness function 
$T(b)=\int_{-\infty}^{\infty}dz\,\rho_A(b,z)$ is the
integral of
nuclear density over longitudinal
coordinate $z$ and depends on the
impact parameter $b$.

The color dipole cross section $\sigma(r)$
introduced
in
\cite{zkl} vanishes like $r^2$ at small $r
\to
0$
due to color screening. This is the heart of the
phenomenon
called
nowadays color
transparency
\cite{bm,zkl,bbgg}.
For this reason nuclear shadowing in (\ref{1.2}) is
dominated
by large size fluctuations corresponding
to
highly
asymmetric sharing of the longitudinal momentum carried by
the
$q$
and $\bar q$ \cite{bk,fs,nz91,kp}. This leads to
$Q^2$
scaling
of
shadowing.

Note that the averaging of the whole exponential
in
(\ref{1.2})
makes this expression different from the
Glauber
eikonal
approximation where $\sigma(r)$ is averaged in
the
exponent.
The difference is known as Gribov's inelastic
corrections
\cite{gribov}.
In the case of DIS the Glauber approximation
does
not
make sense, and the whole cross section is due
to
the
inelastic
shadowing.

For the other case, $t_c \sim R_A$, one has to take into account the
variation
of
$r$ during the propagation of the $q\bar q$
fluctuation
through
the nucleus. At present this can only be done for the
double
scattering
term \cite{kp} in the expansion of the
exponential
in
(\ref{1.2}),
\beq
\frac{\sigma^{\gamma^*A}_{tot}}
{\sigma^{\gamma^*N}_{tot}}
\approx 1
-
\frac{1}{4}\,
\frac{\la\sigma^2(r)\ra}
{\la\sigma(r)\ra}\,
\la T\ra\,
\int
d^2b\,
F_A^2(q,b)\
+\
.\,.\,.
\
,
\label{1.3}
\eeq
\noi
or in
hadronic
representation \cite{kk},
\beq
\frac{\sigma^{\gamma^*A}_{tot}}
{\sigma^{\gamma^*N}_{tot}}
\approx 1
-
\frac{1}{4\pi\sigma^{\gamma^*N}_{tot}}\,
\la
T\ra\,\int
d^2b
\int
dM^2\,\left.\frac{d\sigma(\gamma^*N\to
XN)}
{dM^2\,dt}\right|_{t=0}F^2_A(q,b)
+\
.\,.\,.
\
,
\label{1.3a}
\eeq
\noi
where the mean nuclear thickness and the
formfactor
read,
\beq
\la
T\ra
=
{1\over
A}\,
\int
d^2b\,T^2(b)\
,
\label{1.4}
\eeq
\beq
F_A(q,b)=
\frac{1}{\la
T\ra}\,
\int\limits_{-\infty}^{\infty}
dz\,\rho_A(b,z)\,e^{iqz}\ ,
\label{1.5}
\eeq
\noi
with longitudinal momentum transfer $q=1/t_c$ 
given
by
(\ref{1.1}). In the case of (\ref{1.3}) the uncertain
fluctuation mass is fixed at $M^2=Q^2$, {\it} $q=2m_Nx$.
Two expressions (\ref{1.3})
and
(\ref{1.3a})
are related since the integrated
forward
diffractive
dissociation cross section $\gamma^*N\to XN$
equals
to
$\la\sigma^2\ra/16\pi$.

There are two problems remaining which are under discussion:
\begin{itemize}
\item How the nuclear formfactor can be included 
in the higher order scattering terms
which are of great importance
for
heavy
nuclei?  For instance, the shadowing term in
(\ref{1.3}), (\ref{1.3a})
for
lead is of the order of one at low $x$, so the need of the
higher
order
terms
is obvious.

\item
Even for the double scattering
term
in
(\ref{1.3}) it is still unclear which argument should 
enter the
formfactor.  Indeed, the effective mass of the
$q\bar
q$
fluctuation needed for the coherence time
in
(\ref{1.1})
cannot be defined in the quark representation
with
a
definite
$q\bar q$ separation. On the other
hand,
Eq.~(\ref{1.3a})
exhibits an explicit dependence on $M_X$
and
the
longitudinal momentum transfer is known. However,
unknown
in this case is
the absorptive cross section of the
intermediate
state
$X$.

\end{itemize}

We suggest a solution of both problems in
the
next
section.
The goal of this paper is restricted
to
the study of the difference between the predictions
of the correct
quantum-mechanical
treatment of nuclear shadowing and known
approximations.
We do it on an example of the valence $q\bar q$ part of
the
photon and neglect the higher Fock components containing
gluons
and sea quarks, which may be important if to compare with
data
especially at very low $x$. Nuclear anti-shadowing effect is
omitted
as well, since we believe it is beyond the shadowing
dynamics
({\it e.g.} bound nucleon swelling). Numerical results and a
comparison
with the standard approach are presented in section
3.

\bigskip

\noi
{\large\bf
2. The Green function of a \boldmath$q\bar q$
pair
in nuclear 
medium}
\medskip

We start with the 
generalizing of 
eq.~(\ref{1.2})
for the
case
$l_c\leq
R_A$,
 \beq
 \sigma^{\gamma^*A}_{tot}(x,Q^2)
 =
 \int
 d^2b\,\int\limits_0^1
 d\alpha\,
 \sigma^{\gamma^*A}_{tot}(x,Q^2;b,\alpha)\
 ,
\label{2.1}
\eeq
 \noi
 where
 \beqn
 \sigma^{\gamma^*A}_{tot}(x,Q^2;b,\alpha)
 &=&
 T(b)\,
 \int
 d^2r\left|\Psi_{\gamma^*}(\vec r,\alpha)\right|^2
 \sigma(r)\nonumber\\
 &-&
 2\,{\rm
 Re}\int\limits_{-\infty}^{\infty}
 dz_1\,\rho_A(b,z_1)\,
 \int\limits_{z_1}^{\infty}dz_2\,\rho_A(b,z_2)\,
 A(z_1,z_2,\alpha)\
 .
\label{2.2}
\eeqn
 \noi
 The first term in r.h.s. of (\ref{2.2}) corresponds
 to
 the
 second,
 lowest order
 in
 $\sigma(r)T(b)$,
 term in expansion
 of
 the
 exponential in (\ref{1.2}). The
shadowing terms
are contained in the
 second term
 in
 (\ref{2.2}).
 $\Psi_{\gamma^*}(\vec r,\alpha)$ is
 the
 (non-normalized)
 wave function
 of
 the
 $q\bar q$ fluctuation of the
 virtual
 photon,
 where $\alpha$ is
 the
 fraction
 of the light-cone momentum of
 the
 photon
 carried by the
 quark. An
 explicit
 expression of
 transverse
 and
 longitudinally polarized photons can be
 found
 in
 \cite{gev,nz91}.
 
 The
 function
 $A(z_1,z_2,\alpha)$ in
 (\ref{2.2})
 reads,
 \beq
 A(z_1,z_2,\alpha)
 =
 {1\over
 4}\,\int
 d^2r_1\,d^2r_2\,
 \Psi^*_{\gamma^*}(\vec r_2,\alpha)\,
 W(\vec
 r_2,z_2;\vec
 r_1,z_1)\,
 \Psi_{\gamma^*}(\vec r_1,\alpha)\,
 \sigma(r_2)\,\sigma(r_1)\,
 e^{iq_{min}(z_2-z_1)}\
 ,
\label{2.3}
\eeq
 \noi
 with
 \beq
 q_{min}
 =
 \frac{Q^2\alpha(1-\alpha)
 +m_q^2}{2\nu\alpha(1-\alpha)}.
\label{2.4}
\eeq
\noi
This expression was first suggested in unpublished paper \cite{z}.

The second (shadowing) term in (\ref{2.2}) is
illustrated
in fig.~1.  
\begin{figure}[tbh]
\includegraphics{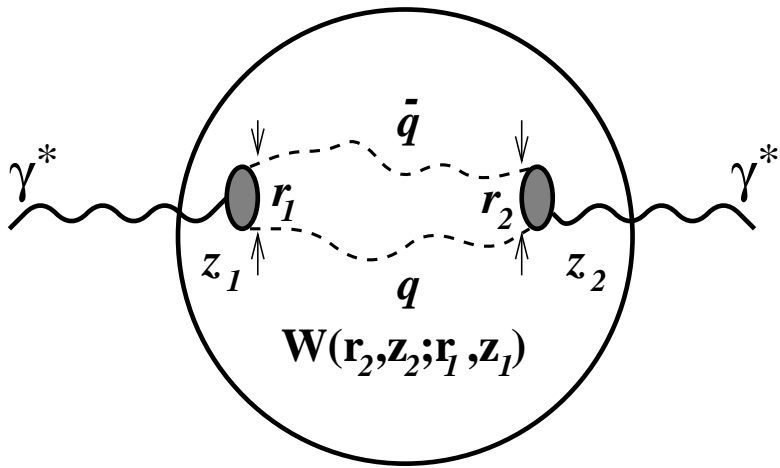}
\begin{center}
\vspace{6.5
cm}
\parbox{13cm}
{\caption[Delta]
{\sl A cartoon for the shadowing (negative)
term
in (\ref{2.2}). The Green function $W(\vec
r_2,
z_2;\vec r_1,z_1)$
results
from the summation over
different
paths of the $q\bar q$ pair propagation
through
the
nucleus.}}
\end{center}
\end{figure}
At the point $z_1$
the
photon
diffractively produces the $q\bar q$
pair
($\gamma^*N\to
q\bar qN$) with transverse separation $\vec r_1$.
The
pair
propagates through the nucleus along
arbitrarily
curved
trajectories (should be summed over) and arrives at
the
point
$z_2$ with a separation $\vec r_2$.  The initial
and
the
final separations are controlled by
the
distribution
amplitude $\Psi_{\gamma^*}(\vec r)$.  While passing
the
nucleus
the $q\bar q$ pair interacts with bound nucleons
via the
cross section $\sigma(r)$ which depends on
the
local
separation $\vec r$.  The function
$W(\vec
r_2,z_2;\vec
r_1,z_1)$ describing the propagation of the pair from
$z_1$
to
$z_2$ also includes that part of the phase shift
between
the
initial and the final photons, which is due
to
transverse
motion of the quarks, while longitudinal motion
is
already
included
in
(\ref{2.4}) via the exponential.

Thus, Eq.~(\ref{2.2}) does not suffer from either
of the two
problems of the approximations (\ref{1.3})
-
(\ref{1.3a}).
The longitudinal momentum transfer is known and
all
the
multiple interactions
are
included.

The propagation function $W(\vec r_2,z_2;\vec
r_1,z_1)$
in
(\ref{2.3}) satisfies
the
equation \cite{z},
 \beqn
 i\,\frac{{\partial}W(\vec
 r_2,z_2;\vec
 r_1,z_1)}
 {{\partial}z_2}
 &=&
 -
 \frac{\Delta(r_2)}{2\nu\alpha(1-\alpha)}\,
 W(\vec
 r_2,z_2;\vec
 r_1,z_1)\nonumber\\
 &-&
 {i\over
 2}\,\sigma(r_2)\,
 \rho_A(b,z_2)\,
 W(\vec r_2,z_2;\vec
 r_1,z_1)\
 ,
\label{2.5}
\eeqn
\noi
with the boundary condition $W(\vec r_2,z_1;\vec r_1,z_1)
=
\delta(\vec r_2-\vec r_1)$.  The Laplacian $\Delta(r_2)$ acts
on
the coordinate $\vec r_2$.  The full derivation
of
(\ref{2.5}) will be given elsewhere.  Here we only
notice
that it looks natural like Schr\"odinger equation with
the
kinetic term $\Delta/[2\nu\alpha(1-\alpha)]$ which
takes
care of the varying effective mass of the $q\bar q$ pair
and
provides a proper phase shift, and $z_2$ plays the role of the time.
The imaginary part of the optical potential describes the absorptive process.

In the ``frozen'' limit $\nu \to\infty$ the kinetic term
in
(\ref{2.5}) can be neglected and 
 \beq
 W(\vec r_2,z_2;\vec
 r_1,z_1)
 =
 \delta(\vec
 r_2-\vec
 r_1)\,
 \exp\left[-{1\over
 2}\sigma(r_2)\,
 \int\limits_{z_1}^{z_2}dz\,\rho_A(b,z)\right]\
 .
\label{2.6}
\eeq
 \noi
When this expression is
substituted into
 (\ref{2.2}) - (\ref{2.3}) and with $q_{min}\to
 0$ one arrives at
 result
 (\ref{1.2})
 with
 $G_{\gamma^*}(Q^2,r)=\int_{0}^{1}d\alpha\,
 \left|\Psi_{\gamma^*}(\vec r,\alpha)\right|^2$.
 
 We can also recover
 the
 approximation
 (\ref{1.3}) - (\ref{1.3a})
if one neglects the absorption of the $q\bar q$  pair 
 in
 the
 medium. Then $W$ becomes the 
 Green
 function of a
 free
 motion,
 \beq\label{free}
 \left.W(\vec
 r_2,z_2;\vec
 r_1,z_1)
 \right|_{\sigma\to
 0}
 =
 \frac{1}{2\pi}\,\int
 d^2k\,
 \exp\left[i\vec
 k(\vec
 r_2-\vec
 r_1)
 +
 \frac{ik^2(z_2-z_1)}
 {2\nu\alpha(1-\alpha)}
 \right]\
 ,
\label{2.7}
\eeq
 \noi
 where $\vec k$ is the transverse momentum of
 the
 quark.
 
 With this expression the shadowing term
 in
 (\ref{2.2})
 reproduces
 the
 second
 term
 in
 (\ref{1.3a}). Indeed,
 the amplitude of the
 photon
 diffractive
 dissociation
 in
 the
 plane
 wave
 approximation
 reads,
 \beq
 f_{dd}(k)
 =
 {1\over
 2}\,\int
 d^2r\,
 \Psi_{\gamma^*}(\vec r,\alpha)\,
 \sigma(r)\,e^{i\vec k\vec
 r}\
 .
\label{2.8}
\eeq
\noi
Therefore, (\ref{2.3}) can be
represented
as,
 \beq
A(z_1,z_2,\alpha)
=
\frac{1}{2\pi}\,
\int
d^2k\,
\left|f_{dd}(k)\right|^2\,
\exp\left[(z_2-z_1)\,
\frac{Q^2\alpha(1-\alpha)+m_q^2+k^2}
{2\nu\alpha(1-\alpha)}\right]
\label{2.9}
\eeq
 \noi
Taking into
account
that
$M_X^2=(m_q^2+k^2)/\alpha(1-\alpha)$
is
the
effective mass squared
of
the
$q\bar q$ pair and substituting
(\ref{2.9})
to
(\ref{2.2}) we arrive
at
eq.~(\ref{1.3a}).

\bigskip

\noi
{\large\bf
3. Numerical
results}
\medskip

We calculate nuclear shadowing for calcium and lead from
the
above displayed equations.  As was mentioned in
the
Introduction, only the valence $q\bar q$-part of the
photon
is taken into account, but the higher Fock
components
containing gluons and sea quarks are neglected, as well
as
the effect of anti-shadowing.  Therefore, we do not
compare
our results with data, but only to the standard
approach
(\ref{1.3}) -
(\ref{1.3a}).

We do the same calculations again, using the free
Green
function (\ref{free}). This makes it possible to
disentangle between
the influence of higher scattering terms and the
formfactor.

We approximate the cross section by
the dipole form $\sigma(r)=Cr^2$, $C\approx 3$, which
is a good approximation at $r > 0.2 - 0.3\ fm$ \cite{kz}.
However, we calculated the proton structure function
$F_2(x,Q^2)$ perturbatively (we fixed the quark masses at 
$m_q=0.3\,GeV$, $m_s=0.45\,GeV$ and $m_c=1.5\,GeV$)
what leads to an additional logarithmic
$r$-dependence at small $r$.
This is important since results in the double-log $Q^2$
dependence of $F_2$. Nuclear shadowing, however, is
dominated by soft fluctuations with large separation
\cite{kp}, therefore, the dipole form of the cross section
is sufficiently accurate.

We use a uniform density for all nuclei, $\rho_A=0.16
\;fm^{-3}$, what is sufficient for our purpose, comparison 
with the standard approach calculated under the same assumption.

Within these approximations it is possible to
solve
(\ref{2.5}) analytically.  The solution is the
harmonic
oscillator Green function with a complex
frequency
\cite{kz}, 
\begin{equation}\label{erg} W\left(\vec
r_2,z_2;\vec
r_1,z_1\right)
=\frac{a}{2\pi\sinh\left(\omega\Delta
z\right)}
\exp\left\{-\frac{a}{2}\left[\left(r_2^2+r_1^2\right)
\coth\left(\omega\Delta z\right)-\frac{2\vec
r_2\cdot\vec
r_1} {\sinh\left(\omega\Delta z\right)} \right]\right\}\
,
\end{equation} 
\noi where 
\begin{eqnarray}
\nonumber\Delta z& = & z_2-z_1 \\ \nonumber\omega^2 & =
&
i\;\frac{C\rho_A}{\nu\alpha\left(1-\alpha\right)}\\ a^2 &
=
& -i\;C\rho_A\nu\alpha\left(1-\alpha\right).
\end{eqnarray}
\noi 
This formal solution properly accounts for all
multiple scatterings and finite lifetime of
hadronic
fluctuations of the photon, as well as for fluctuations
of
the transverse separation of the $q\bar q$
pair.

\begin{figure}[tbh]
\includegraphics{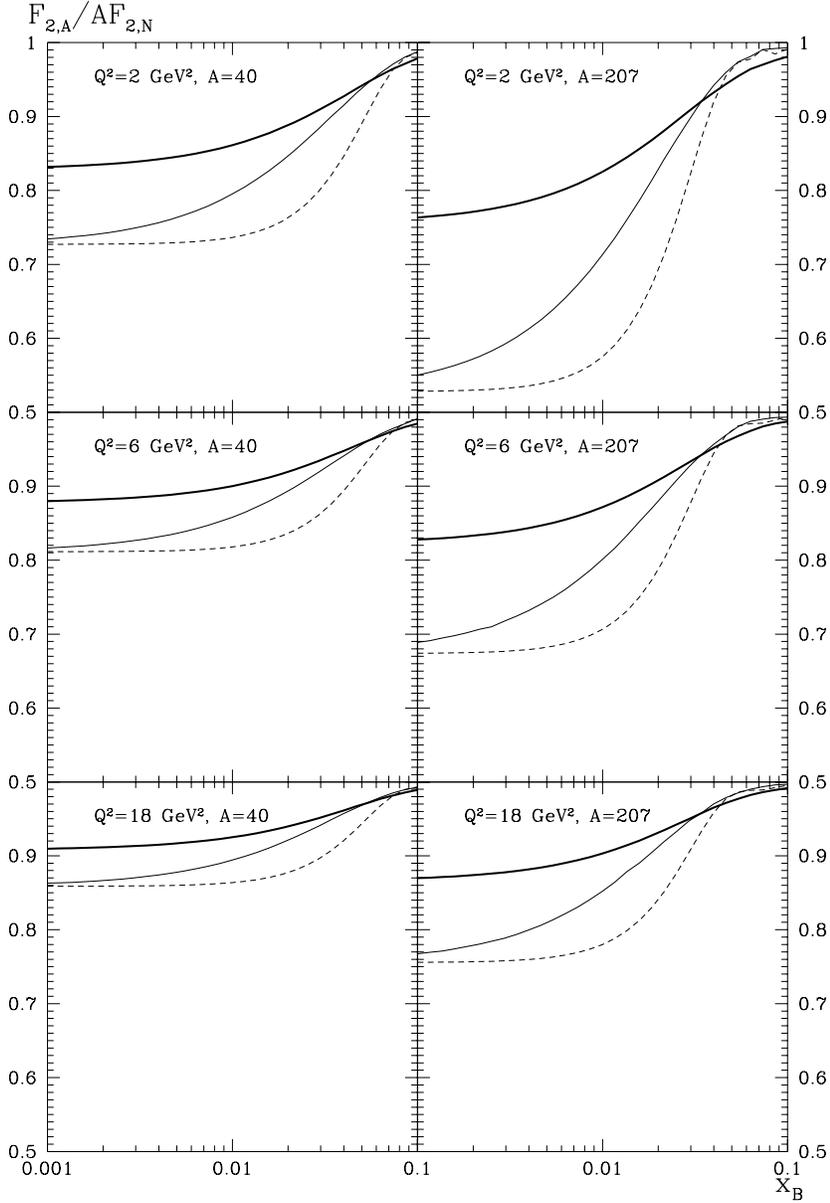}
\begin{center}
\vspace{16cm}
\parbox{16cm}
{\caption[Delta]{
\sl Nuclear shadowing for calcium and lead. The
dotted
curve is calculated in the standard approach
(\ref{1.3}).
The thin solid curve
corresponds
to the double scattering approximation with the free Green
function,
(\ref{free}), and the thick
solid
curve shows the full calculation,
(\ref{erg}).}
\label{bild}}
\end{center}
\end{figure}

The results of calculations are shown in
fig.~\ref{bild}.
The dashed curves show predictions of (\ref{1.3}) which
we
call standard approach.  The mean values of $\sigma^2$
and
$\sigma$ are calculated using the same $q\bar
q$
distribution functions \cite{gev,nz91} as in (\ref{2.3}) and
the
intermediate state mass is fixed at $M^2=Q^2$. At low $x
<
0.01$ shadowing saturates because $q=2m_Nx \ll
1/R_A$. The
thin solid curve also corresponds to a double
scattering
approximation, {\it i.e.} absorption (the second term
in
(\ref{2.5})) is omitted.  However, the formfactor is
treated
properly, {\it i.e.} the kinetic term in (\ref{2.5})
taking
into account the relative transverse motion of the $q\bar
q$
pair, correctly reproduces the phase shift.  The
difference
between the curves is substantial.  The thin solid
curve
does not show saturation even at
$x=0.001$.

The next step is to do the full calculations and
study
importance of the higher order rescattering terms
in
(\ref{2.5}).  The results are shown by the thick
solid
curves.  Higher order scattering brings another
substantial
deviation (especially for lead) from the standard
approach.
At very low $x$ the curves saturate at the level given
by
(\ref{1.2}).

\bigskip

\noi
{\large\bf
4. Conclusions and
outlook}
\medskip

We suggest a solution for the problem of nuclear
shadowing
in DIS with correct quantum-mechanical treatment of
multiple
interaction of the virtual photon fluctuations and
of the nuclear
formfactor.  We perform numerical calculations for 
$q\bar
q$ fluctuations of the photon and find a
significant
difference with known approximations.
Realistic
calculations to be compared with data on nuclear
shadowing
should incorporate the higher Fock components which
include
gluons.  The same path integral technique can be applied in this
case.
The $x$-dependence of the dipole cross section
$\sigma(r,x)$
(correlated with $\vec r$ \cite {kp98}) should be taken
into
account.  One should also include the effect
of
anti-shadowing, although it is only a few percent.
A
realistic form for the nuclear density should be used (this
can
be done replacing $\rho_A(b,z)$ by a multistep function
like
in \cite{kz}). We are going to settle these problems in
a
forthcoming
paper.

{\bf Acknowledgements:} We are grateful for stimulating
discussions to J\"org
H\"ufner and Gerry
Garvey who read the paper and made many useful
comments.

The work of J.R. and A.V.T was supported by the Gesellschaft f\"ur 
Schwerionenforschung, GSI, grant HD H\"UF T,
and B.K. was partially supported by European Network:
Hadronic Physics with Electromagnetic Probes, No FMRX CT96-0008,
and by INTAS grant No 93-0239ext.
J.R and A.V.T. greatly acknowledge the hospitality at the MPI.


\begin{thebibliography}{MMM}

\bibitem{kancheli}
O.V.~Kancheli,
Sov.  Phys.  JETP Lett.  {\bf
18}
(1973)
274

\bibitem{glr} L.V.~Gribov,
E.M.~Levin
and
M.G.~Ryskin, Phys.
Rept.
{\bf
100}
(1983)
1;

\bibitem{mq}
A.H.~Mueller and J.~Qiu, Nucl. Phys.
{\bf
B268},
(1986)
427
\bibitem{q} J.~Qiu, Nucl. Phys. {\bf
B291}
(1987)
746

\bibitem{lp} L.D.~Landau and
I.Ya.~Pomeranchuk, Dokl. Akad. Nauk
SSSR, {\bf 92} (1953) 535; {\it ibid} 735.
In English see in L.D.~Landau {\it The Collected 
Papers of L.D.~Landau} (Pergamon Press, New York, 1965)

\bibitem{bauer}
T.H.~Bauer,
R.D.~Spital,
D.R.~Yennie
and
F.M.~Pipkin,
Rev. Mod. Phys. {\bf
50}
(1978)
261

\bibitem{fs}
L.L.~Frankfurt
and
M.I.~Strikman, Phys.  Rept.
{\bf
160}
(1988)
235

\bibitem{bl} S.J.~Brodsky and H.J.~Lu,
Phys. Rev. Lett. {\bf 64} (1990) 1342

\bibitem{nz91} N.N.~Nikolaev and
B.G.~Zakharov,
Z.
Phys. {\bf
C49}
(1991)
607

\bibitem{mt} W.~Melnitchouk and A.W.~Thomas,
Phys. Lett. {\bf B317} (1993) 437

\bibitem{weise} G.~Piller, W.~Ratzka and W.~Weise,
Z.Phys.{\bf A352} (1995) 427

\bibitem{kp} B.Z.~Kopeliovich
and
B.~Povh,
Phys.Lett. {\bf B367} (1996) 329; Z.Phys. {\bf A356}
(1997)
467

\bibitem{zkl}
Al.B.~Zamolodchikov,
B.Z.~Kopeliovich
and
L.I.~Lapidus,
Sov.
Phys. JETP Lett.
{\bf
33},
(1981)
612

\bibitem{bm} S.J.~Brodsky and A.~Mueller,
Phys. Lett. {\bf
B206}
(1988)
685

\bibitem{bbgg} G. Bertsch, S.J. Brodsky, A.S. Goldhaber
and
J.F. Gunion,
Phys. Rev. Lett. {\bf 47},
297
(1981)

\bibitem{bk}
J.D.~Bjorken and J.~Kogut, Phys.
Rev. {\bf
D8}
(1973)
1341

\bibitem{gribov}
V.N.~Gribov,
Sov.  Phys.  JETP {\bf 57}
(1969)
1306

\bibitem{kk} V.~Karmanov and L.~Kondratyuk,
JETP
Lett.,
{\bf 18}
(1973)
451

\bibitem{gev} S.~Gevorkyan, A.M.~Kotsinian and V.M.~Jaloian,
Phys.Lett.{\bf 212B} (1988) 251

\bibitem{z} B.G.~Zakharov, 'Light-cone path integral approach
to the LPM effect', MPI-H-V44-1997 (unpublished).

\bibitem{kz} B.Z.~Kopeliovich
and B.G.~Zakharov,
Phys.Rev. {\bf D44} (1991)3466.

\bibitem{kp98} B.Z.~Kopeliovich and B.~Povh,
hep-ph/9806284

\end{thebibliography}
\end{document}